# TITLE:
Fabrication and characterization of high-Q silicon nitride membrane resonators

## AUTHORS AND AFFILIATIONS:
Atkin D. Hyatt[1], Oscar A. Flores[1], Aman R. Agrawal[1], Charles A. Condos[1], Dalziel J. Wilson[1]

[1]Wyant College of Optical Sciences, University of Arizona, Tucson, Arizona 85721, USA

atkindavidhyatt@arizona.edu (corresponding author)
oscaraf@arizona.edu
amanagrawal@arizona.edu
cacondos@arizona.edu
dalziel@arizona.com

## SUMMARY:
We trace the life cycle of a silicon nitride membrane resonator, from its design to its fabrication and, ultimately, its characterization.

## ABSTRACT:
Silicon nitride membranes are a powerful and ubiquitous optomechanical resonator technology, enabling high mechanical Q, low optical loss, and enhanced optomechanical coupling via a panoply of strain-, phononic-, and photonic-crystal engineering techniques. Fabrication and characterization of silicon nitride membranes has become a form of tacit knowledge in optomechanics research groups. Here we present a video run-through of the design, fabrication, and characterization of a contemporary silicon nitride membrane resonator (specifically, a centimeter-scale $Si_3N_4$ nanoribbon supporting $Q>10^8$ torsion modes). Our tutorial can serve as a starting point or refresher for practitioners in the field.

## INTRODUCTION:
Silicon nitride membranes have played an important role in quantum optomechanics [1] over the last two decades, beginning with the discovery that a commercial membrane (a TEM slide) can support drum modes with Q-frequency products exceeding $10^{13}$ Hz and be coupled to a high finesse optical cavity using the "membrane-in-the-middle approach" [2,3,4]. Gradually it was appreciated how tensile stress and modeshape affect the mechanical Q (through an effect called "dissipation dilution" [5,6]), spurring the invention of micropatterned membranes—trampoline [7,8], 2D an1D phononic crystal [9,10], fractal [11], and perimeter mode resonators [12]—with Q factors as high as $10^9$ [6]. In conjunction with cryogenics, these devices have enabled landmark experiments such as ground state cooling [13,14], ponderomotive squeezing [15], optomechanical entanglement [16], and displacement measurement beyond the Standard Quantum Limit [17,18]. They also feature prominently in proposals for next generation optomechanical technologies and probes for new physics, including membrane-based force microscopes [19], accelerometers [20], spectroscopes [21], quantum memories [22], microwave-to-optical photon transducers [23], and dark matter detectors [24].

Despite their popularity, design, fabrication, and characterization of silicon nitride membrane resonators remains a niche discipline within the optomechanics community, relying on bespoke techniques handed down by word of mouth and dissertations [25-32]. A recent review of interferometric characterization of nanomechanical resonators demystifies the latter task [33], and modeling of dissipation dilution has become simple with commercial finite element simulation software [28]. Nanofabrication, however, remains an obstacle to entry, since the procedure, while straightforward, involves a sequence of delicate steps whose nuances are often left out of traditional verbal descriptions and process flow diagrams.

In this publication, we outline the design, fabrication, and characterization of a silicon nitride membrane resonator in video format. We use a centimeter-scale ribbon resonator [34]—a simple yet robust design gaining popularity for torsional quantum optomechanics [35,36] and precision measurement experiments [37,38]—as an example. However, the procedure is easily adaptable to more complex resonator geometries. We start with a brief overview of the simulation of membrane resonators in COMSOL Multiphysics before pivoting to fabrication, which is our focus. We conclude with discussion of characterization using an optical lever. Our tutorial can serve as a starting point or refresher for practitioners in the field.

**PROTOCOL:**
1. Simulation and design [32,28]
   *Choose resonator parameters to optimally fit a predetermined objective*
   1.1. Build a COMSOL model of the generic resonator geometry of interest.
      1.1.1. Start the COMSOL model wizard and create a 2D simulation. Under the structural mechanics physics menu, select either "plate" or "shell."
         NOTE: the difference between plate and shell is minimal in this context.
      1.1.2. For the purposes of design, we are interested in finding the frequency, mode shape, effective mass, and quality factor of an oscillator's modes. Thus, we select "Eigenfrequency, Prestressed" in the "Select Study" menu.
      1.1.3. Construct the resonator geometry and set the material of all domains to stoichiometric silicon nitride (Listed as "$Si_3N_4$ – silicon nitride" in the COMSOL material library).
         NOTE: The mechanical properties of silicon nitride vary depending on stoichiometry and deposition process. We alter the material density to 2700 kg/m$^3$ to reflect the films used in this work (see below).
      1.1.4. Add an initial stress and strain to the model corresponding to the prestress of the $Si_3N_4$ film. (In our case, we specify 1 GPa.) Then add a fixed constraint node and select the device's clamps.
      1.1.5. Under the "Mesh" node, choose the "extremely fine" element size option. We find that choosing this option is important for accurate dissipation dilution simulation. To further increase accuracy, the mesh density can be scaled in the "Mesh Density" window.
      1.1.6. Go to the stationary study step and ensure that the "Include geometric nonlinearity" box is checked. Under the eigenfrequency study step, choose how many modes to solve for and run the simulation.

- 1.1.7. Under the results node, add a global evaluation which estimates the resonator's quality factor arising from dissipation dilution—the ratio of the total kinetic and elastic strain energy times the intrinsic quality factor [28,32,39].
- 1.1.8. Modify the resonator's geometry until the effective mass, frequency, and quality factor meet specifications.
1.2. GDSII CAD
- 1.2.1. Transfer the design geometries to a GDSII CAD file and populate a wafer CAD. In this work, we use positive photoresist so only the area around the resonator is modeled.
- 1.2.2. In addition to a layer of resonators, we add a bottom layer representing back windows on each device to remove material from behind each device. Doing so also hastens the wet etch process later in the protocol.
- 1.2.3. Finally, dice lines and four alignment markers are placed on the top, bottom, and sides of the wafer CAD on both layers. This helps properly align the wafer backside when running the photolithography machine.

2. Wafer-scale fabrication [1]

*Pattern the design(s) found in the previous section onto a double-sided $Si_3N_4$-on-Si wafer. Note that while $Si_3N_4$ deposition can be performed by the reader, our facilities lack the necessary equipment to do so and, as such, start with commercially sourced wafers (WaferPro, for this study). An overview of the fabrication process is shown below in Figure 1.*

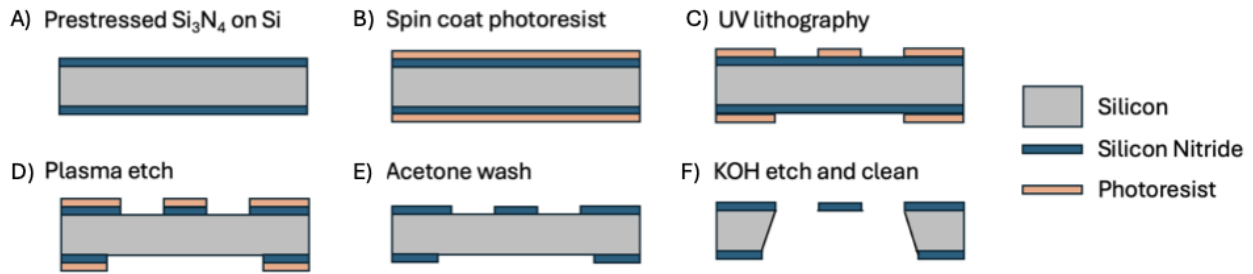

**Figure 1: Fabrication Overview.** First, a) a thin film of silicon nitride is deposited onto a bare silicon substrate. The wafer is then b) coated with photoresist for c) photolithography. The pattern traced out in c) is used as a protective mask for d) reactive ion etching to carve the resonator designs into the film. After dicing, e) acetone is used to remove the resist before f) potassium hydroxide solution removes the silicon substrate.

2.1. Deposit a high stress thin film of $Si_3N_4$ onto a bare silicon substrate using low pressure chemical vapor deposition (LPCVD). While we do not perform this step, typical deposition parameters are listed below:

    Quartz temperature: 800 °C
    Pressure: 10 mTorr
    Injected gas mixture: Dichlorosilane ($SiH_2Cl_2$) and ammonia ($NH_3$)
    **CAUTION**: $SiH_2Cl_2$ is combustible, corrosive, and is acutely poisonous if inhaled and should only be handled with extreme care and proper PPE.

2.2. Hexamethyldisilazane (HMDS) vapor-phase priming [40]
- 2.2.1. A clean, double-sided $Si_3N_4$-on-Si wafer is placed on two glass slides inside a wide glass petri dish. It's important to space the slides such that only the outer edge of the wafer is supported, improving the backside priming in later steps.

2.2.2. A few drops of HMDS are placed along the edge of the petri dish with a disposable capillary pipette. Under a fume hood, the glass petri dish is covered and heated to 90 °C for one minute on a hot plate, evaporating the HMDS and allowing it to adhere to the $Si_3N_4$.
**CAUTION**: HMDS is combustible, an irritant, and causes long-term health problems if inhaled and should only be handled under a fume hood.

2.2.3. After one minute, the petri dish cover is gently removed and any remaining HMDS is vented from the dish. The wafer is then moved to a plastic petri dish with a cleanroom wipe on the bottom.

2.3. Spin coating resist

2.3.1. Before depositing resist, a hotplate is preheated to 115 °C with two clean glass slides on top. This will ensure the hotplate is at the prescribed temperature later in the protocol.

2.3.2. The primed wafer is centered inside a spin coater and held in place with a vacuum chuck. The spin coater is left open for later steps.

2.3.3. Using a needle and syringe, we extract no more than 4 mL of Microposit S1813 photoresist [41]. The syringe is first gently flicked, and a small amount of fluid is dispensed, to remove air bubbles.
**CAUTION**: S1813 is a combustible irritant and should only be handled with proper PPE and away from open flames or sparks.

2.3.4. The needle is then holstered and replaced with a 2 μm filter to remove large particulates. As in the previous step, the syringe is gently flicked and we dispense 3-5 drops of fluid to remove remaining air bubbles in the filter. Before continuing, make sure at least 3 mL of resist remains in the syringe.

2.3.5. The resist is then deposited drop by drop onto the wafer to form one large pool. As this pool gets larger, resist is deposited near its edge to ensure it stays relatively centered on the wafer. If any bubbles appear on the pool surface, use the needle tip to pop them before continuing.

2.3.6. The spin coater lid is then locked and set to spin at 3000 rpm for 30 seconds with an acceleration of 3000 rpm/s. These settings, according to the resist manufacturer, will create a uniform 1.5 μm thick layer across the wafer surface [6].

2.3.7. The finished coat is inspected for bubbles, streaks, and non-uniformity. If any are spotted, the spin coater is shut and the wafer surface washed with acetone and isopropyl alcohol (IPA). Running the spin coater during this step enhances the removal. All steps in 2.3 are repeated to form a new layer.
**CAUTION**: Acetone and IPA are both combustible irritants. They should only be handled with proper PPE, and away from open flames or sparks.

2.3.8. If the resist layer is defect-free, the vacuum chuck is disabled and the wafer is transferred to the preheated glass slides for a one-minute bake [6]. Baking the wafer evaporates leftover solvent and prepares it for photolithography.

2.3.9. After baking, the wafer is removed from the slides. The wafer can proceed to the photolithography step.

2.4. Photolithography

2.4.1. The coated wafer is loaded into a maskless alignment (MLA) photolithography machine. We make sure to center the wafer the best we can to minimize global rotation and offset errors.
2.4.2. The populated wafer file from step 1.2 is loaded onto the computer and converted to a file type which can be understood by the MLA software. We begin by first patterning the topside layer populated with resonator designs.
2.4.3. After loading the wafer and CAD file, we select the corresponding wafer shape and size. The machine used in this work then automatically determines alignment by using an edge-detection algorithm.
2.4.4. After alignment, the option to include global rotation error is selected and the appropriate beam exposure parameters loaded. We determined that a beam intensity of 140 mJ/cm$^2$ at a wavelength of 375 nm produces optimal results for 1.5 µm of our resist [1].
2.4.5. With all exposure parameters loaded and the wafer aligned, we begin the exposure. For a 100 mm wafer divided into 12 mm square chips, this process usually takes about 20 minutes.
2.4.6. Five minutes before the end of the photolithography exposure, two large dishes are rinsed with deionized (DI) water and blow dried with compressed nitrogen gas.
2.4.7. One dish is filled with about 75 mL of MF-319 developer [42] while the other is filled with a large amount of DI water.
**CAUTION**: MF-319 is corrosive, irritating, and poses long-term health risks. Only use with proper PPE.
2.4.8. Once the exposure is complete, the wafer is unloaded from the MLA and transferred to the developer to be left undisturbed for 20 seconds.
2.4.9. After 20 seconds of development, the solution is then agitated to remove exposed resist for an additional 40 seconds by gently swirling the dish in a circular motion. If exposed resist remains, the mixture can be agitated for an additional 30 seconds, but caution must be taken to not overexpose the wafer [42].
2.4.10. The wafer is then transferred to the second dish filled with DI water to dilute off residual developer. It is then gently washed with DI water using either a wash bottle or DI water gun on both sides to remove the rest.
2.4.11. Finally, a compressed nitrogen gas gun pointed along the wafer removes the DI water from the wafer surface. It is important to use low pressure and proper gun orientation to minimize the chance of damage.
2.5. Reactive ion etch [40]
2.5.1. The developed wafer is then loaded into a reactive ion etcher to transfer the resist patterns onto the $Si_3N_4$ film. The following process is ran for 5 seconds at a time, removing 30 nm of $Si_3N_4$ [40]:
Gas composition: 30 sccm of sulfur hexafluoride ($SF_6$) and 10 sccm of Argon (Ar)
High frequency bias: 50W
Power of inductively coupled plasma: 1000 W
Chamber pressure: 10 mTorr

- 2.5.2. We repeatedly run this process to remove about 30 nm of $Si_3N_4$ per process until the exposed film appears silver, an indicator that the substrate is exposed. Finally, the wafer is removed from the reactive ion etcher and is ready for backside patterning.
- 2.6. Backside patterning
    - 2.6.1. The wafer is then placed back into the spin coater with the device side facing down. A 1.5 μm layer of photoresist is deposited identical to step 2.3.
    - 2.6.2. With the new resist layer facing up, the wafer is loaded back into the MLA, rotated 180°. The wafer CAD is reloaded into the MLA software, converted to an acceptable filetype, and mirrored along the respective axis. Viewing the converted CAD ensures that the CAD orientation matches that of the wafer.
    - 2.6.3. Steps 2.4.3 through 2.5.4 are repeated.
- 2.7. Wafer cleaving
    - 2.7.1. Using the dice lines patterned into the silicon nitride as a guide, two glass slides are placed on the wafer such that they are parallel to the substrate crystal plane. A diamond-tipped scribe is carefully inserted between the slides and dragged along the dice lines, scoring along the crystal plane.
    - 2.7.2. The slides are then re-aligned to a new dice line and the scribe used to score along another part of the wafer. We repeat this scoring process until all edges of individual chips have been scribed.
    - 2.7.3. The wafer is then placed on top of the slides such that a dice line is equidistant from both. If the score was deep enough and well aligned, a small amount of force to the top of the wafer is enough cleave the wafer along the crystal plane, resulting in a perfect break. The pieces of the wafer are continuously broken in this fashion to make individual chips.
3. Chip-scale processing [32]

    *Release the patterned resonators from their silicon substrate to produce free-standing films.*
    - 3.1. Preparation
        - 3.1.1. First, 40 mL of 45% concentration potassium hydroxide (KOH) solution is transferred to a vessel (or beaker) and heated to 86 °C using a hotplate. The solution is covered to promote a uniform temperature gradient.
        **CAUTION**: KOH is corrosive and an irritant and should only be handled with proper PPE under a fume hood.
        - 3.1.2. A plastic sample holder (with a lid) is given a strong puff of nitrogen gas to remove dust while a thin metal washer the size of the chip is rinsed with IPA and blow dried with compressed nitrogen gas. The washer will act as a stand for the released membrane inside this holder.
        - 3.1.3. Diced chips are carefully peeled from their tape and dipped in a sonicated acetone bath for at least 30 seconds to remove resist and surface contaminants. The acetone is then washed off with a quick IPA rinse and dried with compressed nitrogen gas.
        - 3.1.4. OPTIONAL: To thin the $Si_3N_4$ film and/or to remove stuck-on contaminants, the chip can be dipped in a 10% hydrofluoric acid (HF) buffer solution to remove 1.55 nm/min. This step is optional in that it can be performed before or after the KOH etch.

**CAUTION**: HF is acutely poisonous, corrosive, and requires specific safety considerations before being used. Only handle with proper training, acid-safe gloves/aprons, and with calcium gluconate gel ready in case of exposure.

3.1.5. To prevent foreign contaminants from reattaching to the chip, it is immediately placed in a custom polytetrafluoroethylene (PTFE) holder [1] under a fume hood. This holder is shown in Figure 2.

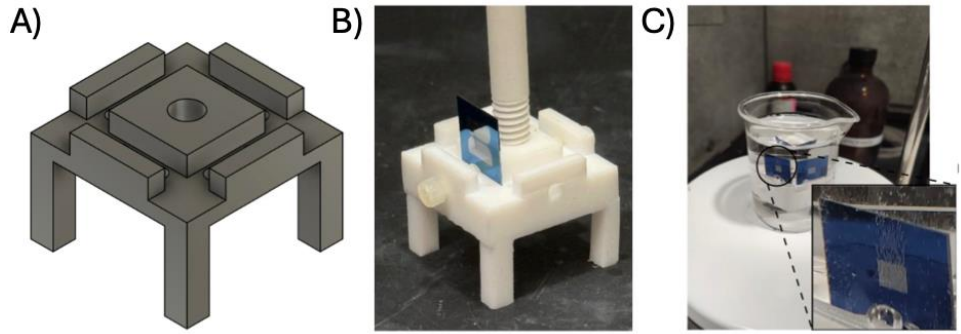

**Figure 2: Custom Chip Holder.** Borrowed with permission from [32]. Our custom chip holder is designed to keep chips upright while maximizing exposed area for KOH etching. a) CAD drawing showing the basic structure of the device. The final structure is machined out of a small block of PTFE plastic for chemical resistance. b,c) Chips are placed upright in the holder and a PTFE stick is used to lower them into a chemical bath.

3.2. KOH wet etch

3.2.1. The custom chip holder is lowered into the KOH bath. The solution is then covered to maintain a uniform thermal gradient.

3.2.2. The KOH reacts with silicon preferentially along the <100> crystal plane of the substrate. While the etch rate along other directions is small, it remains nonnegligible, etching corner features slowly [2]. Since KOH etching takes about one day, a camera is set up next to the etch for long-distance monitoring.

3.2.3. After all silicon around the resonator has been etched and the film is released, the reaction is stopped by removing the vessel from the hot plate. Due to the viscosity of KOH, removing the device would likely shatter the film [8] so the KOH bath is gradually replaced with DI water by iteratively pipetting out KOH and pipetting in DI water, never letting the liquid level fall below the top of the chip. This process is repeated until twice the vessel volume has been removed.

3.2.4. To prevent cross contamination with KOH and solvents later in this protocol, the chip holder is transferred to a fresh DI water bath.

3.2.4.1. First, a large vat is filled with excess DI water, enough to cover both the etching and DI vessels with room for the holder above. The new DI water bath is placed in the vat using tongs, followed by the etching vessel.

3.2.4.2. The dip stick is used to gently lift the chip holder out of the etching vessel and, while keeping the chip submerged, transferred to the DI bath. Both vessels are removed from the vat with tongs, leaving the chip holder in a clean DI water bath.

3.3. Cleaning the released film

- 3.3.1. After the chip has been transferred to a fresh DI water bath, the DI water is gradually replaced with IPA in the same fashion as step 3.2.5. Again, never letting the liquid level fall below the top edge of the chip.
- 3.3.2. After two baths worth of liquid has been transferred out, the process is repeated, replacing IPA with methanol.
- 3.3.3. After two baths worth of liquid has been transferred out, the device can be removed from the methanol bath and carefully blow dried along the chip plane with a gentle current of nitrogen gas.
- 3.3.4. Inspect the device under a microscope and look for debris, leftover resist, or flaws and clean as necessary by placing the chip back in methanol for some time and re-drying.
- 3.4. HF wet etch and additional cleaning
  - 3.4.1. If the debris remains resilient against the methanol cleaning step 3.3.4, the chip may need to be etched in a 10% HF buffer solution. Note that this reacts with $Si_3N_4$ at a rate of about 1.55 nm/min. In many cases, a quick dip is all that's necessary to strip the outer layer of film.
  - 3.4.2. After dipping in HF, the device is immediately transferred to a DI water bath for 30 seconds to dilute the HF.
  - 3.4.3. The chip is then transferred to an IPA bath for another 30 seconds and finally transferred to a methanol bath for several minutes.
  - 3.4.4. After the desired time has elapsed, the chip is removed from the methanol bath and gently dried with a light breeze of nitrogen gas.
  - 3.4.5. The chip is then re-inspected and re-cleaned as necessary. After the device is deemed clean, it is gently placed inside its holder atop the washer from step 3.1.2.
4. Characterization
   4.1. Characterization begins by loading the released membrane face up into a vacuum chamber with a transmission port for optical probing. To minimize mechanical loss, the chip is simply rested in the chamber, no tape or glue is used to hold it in place [25].
   4.2. Optical lever (OL) setup to measure mode displacement [34,35]. (The OL method has numerous practical advantages; however, since it requires probing angular deflection, it is less sensitive than interferometry for some modeshapes. We refer the reader to [33] for a useful overview of interferometric displacement readout of membrane resonators.)
   - 4.2.1. First, a collimated laser beam is focused onto the sample with a spot size comparable to the largest feature being probed. This is important since the optical lever sensitivity is proportional to both spot size and reflected power [35].
   - 4.2.2. The laser beam is then aligned to be as close to normal incidence onto the sample as possible, typically by retroflection into an optical fiber. (Though the incidence angle doesn't affect the sensitivity much, it simplifies alignment later.)
   - 4.2.3. With the laser at normal incidence, a beamsplitter is inserted between the sample and focusing lens. The beamsplitter picks off the light reflected off the resonator.
   - 4.2.4. A balanced photodetector is placed along the picked-off beam path a good distance from the setup, ideally greater than the Rayleigh length of the focused beam, to maximize sensitivity [35]. A diagram of the OL setup is given in Figure 3a.
   4.3. Displacement readout and thermal noise measurement

4.3.1. Translate the incident beam along the resonator to focus onto a region of high mode curvature. This can be found by referencing the mode shapes found in the simulations in part 1. Realign the rest of the setup as necessary.

4.3.2. With light incident on the split (or quadrant) photodetector, the detector output is sent to a digitizer (National Instruments PXI 4461). The real-time power spectral density (PSD) of the digitized signal is then computed using the Fast Fourier Transform (FFT) method [34].

4.3.3. For a sufficiently sensitive OL measurement, thermal noise peaks appear in the broadband signal PSD. To identify specific modes, we compare the location of thermal noise peaks to the eigenfrequencies predicted by the simulations in Part 1. An example thermal noise peak is shown in Figure 3b. For this measurement, an RMS average of several PSD estimates (periodograms) is taken.

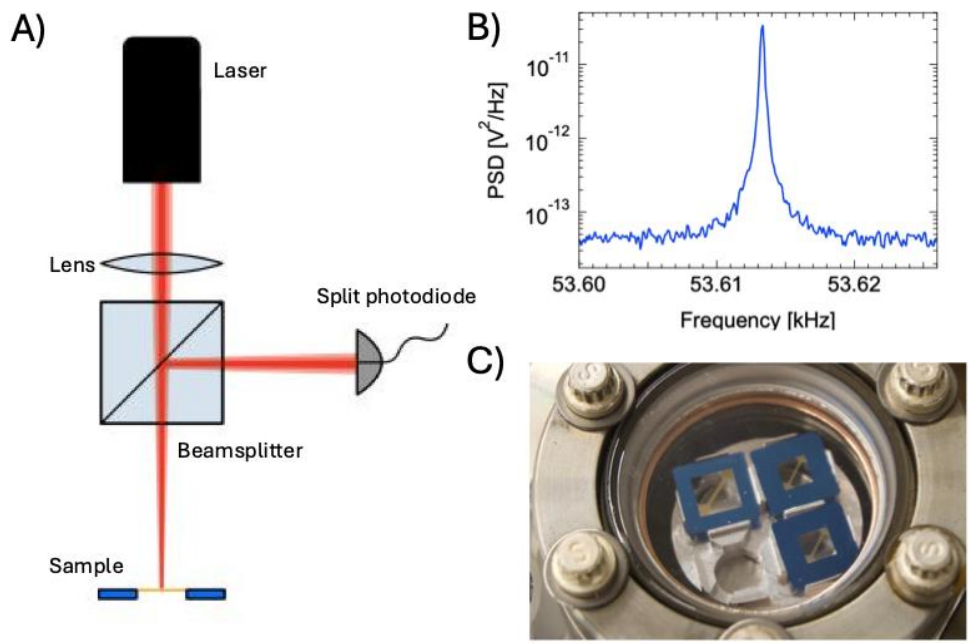

**Figure 3: Characterization Setup.** a) A simple cartoon outlining the basic setup and operation of an optical lever. A collimated laser beam is focused onto the sample using a lens. A beamsplitter picks off the retroflected light, directing it to a split photodiode. b) Example of a thermal noise signal averaged 50 times with a resolution bandwidth of 0.1 Hz. c) Resonators are rested on a custom aluminum rack, preventing extrinsic loss mechanisms from deteriorating the quality factor.

4.4. Energy ringdown

4.4.1. The energy contained within an oscillating mode is proportional to the area under the PSD. We therefore begin by tracking the integral over a narrow band centered at the modal resonant frequency peak. As energy is added to the mode and subsequently dissipates, this will characterize its dissipation.

4.4.2. When energy is added in the form of a coherent drive, the peak (and the area underneath it) will increase. This can be achieved by either placing a piezo actuator on the side of the vacuum chamber or sending a second laser beam to the sample. In either case, the source is modulated at the resonant frequency of the mode.

NOTE: Adding energy to an oscillator can also be done by giving it a "kick". An impulse force from tapping the vacuum chamber creates an instantaneous, incoherent white drive which excites many modes at once at the cost of inconsistency.

4.4.3. After the drive is tuned off, the mode oscillation energy will decay exponentially. The damping rate of the oscillator is inferred from a fit performed after the "ringdown". The modal Q is calculated by dividing the resonant frequency by the damping rate.

4.4.4. This ringdown procedure is repeated several times to find an average Q and to verify the consistency of the results.

4.5. Stroboscopic readout

4.5.1. If the device appears to ringdown in a nonlinear fashion or the results vary wildly between ringdowns, the device may be interacting with the probe and driven via photothermal processes [43]. To avoid this issue, we employ a stroboscopic ringdown [10] in which the probe is shuttered on for a few seconds and shuttered off for several more. By only exposing the probe for a few seconds at a time, heating of the resonator is minimized.

**REPRESENTATIVE RESULTS:**

We demonstrate the protocol by fabricating a wafer consisting of several 100 nm thick diagonal ribbon resonators [34] and characterizing their first few vibrational modes. In this work, we let the ribbon be 7 mm long and 400 μm wide with 662 μm diameter fillets. We note that while the optical lever used in this work is naturally suited to torsional modes, it can also detect transverse flexural motion by positioning the beam at a location of non-zero angular detection. As an illustration, we use the OL to measure both flexural and torsional modes of our ribbons.

We start by simulating the first two modes of this design in COMSOL. The eigenfrequency analysis yields two primary modes: a 53 kHz flexural mode with a Q of 40 million and a 70 kHz torsional mode with a Q of 200 million. Looking at their mode shapes gives intuition on how to probe them with the optical lever. As seen in Figure 4c, the flexural mode has the largest angular displacement halfway between the center of the ribbon and its fillets while the torsion mode has the greatest deflection at the ribbon's center. We note this for later when we attempt characterization.

Next, we create a GDSII CAD of a 100 mm circular wafer and divide it into 37 total 12 mm$^2$ chips. After adding the appropriate alignment markers and dicing lines, we populate it with several designs. While this publication will focus on just the ribbons, we include multiple geometries to highlight the flexibility of the protocol. The populated wafer CAD is shown in Figure 4a and 4b.

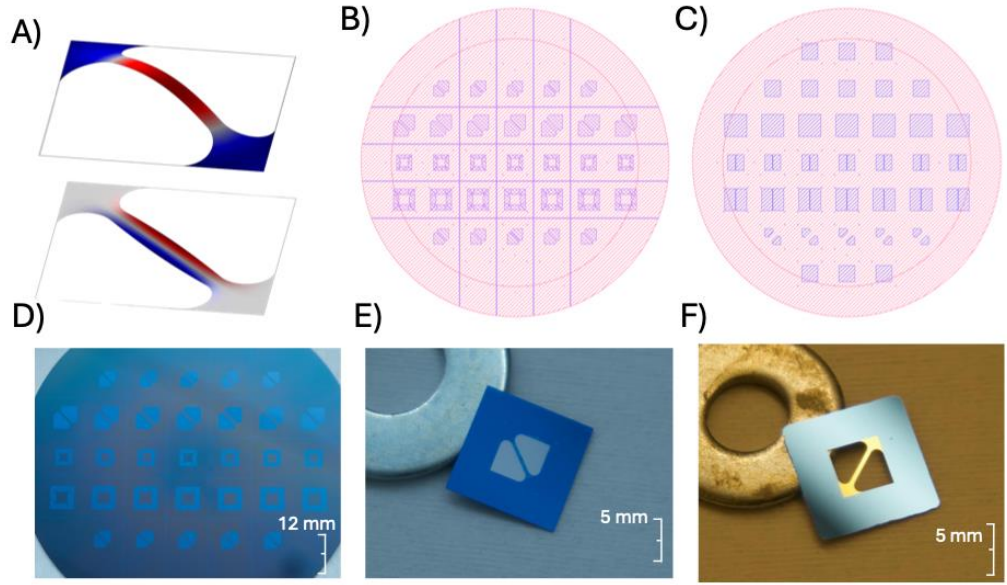

**Figure 4: Nanoribbon Design and Fabrication.** a) Following the protocol, we start by simulating the behavior of the ribbon listed in the main text, noting the shape of the flexural mode (top) and torsional mode (bottom). b) We populate a GDSII CAD file with these ribbons and other resonator designs. Though we focus on a ribbon, other designs are highlighted to emphasize the versatility of our method. c) Back windows are added to release silicon from beneath the top film. d) Photolithography creates the device patterns as a mask for reactive ion etching. e) The exposed top film is blasted away by plasma and the wafer is cleaved into individual chips. f) Finally, KOH removes the substrate and gradual dilution allows for the safe removal from a liquid bath.

Following the steps laid out in sections 2 and 3 of the protocol, we release fabricate the designed wafer and release several ribbons. Based on our understanding of the mode shapes from earlier, we align the optical lever roughly to one the areas of maximum angular deflection at a time and record their thermal motion on a split photodiode. After confirming that the mode frequencies match their predicted values, we perform an energy ringdown by incoherently driving them and tracking the energy dissipation as a function of time. The results of these measurements are given in Figure 5.

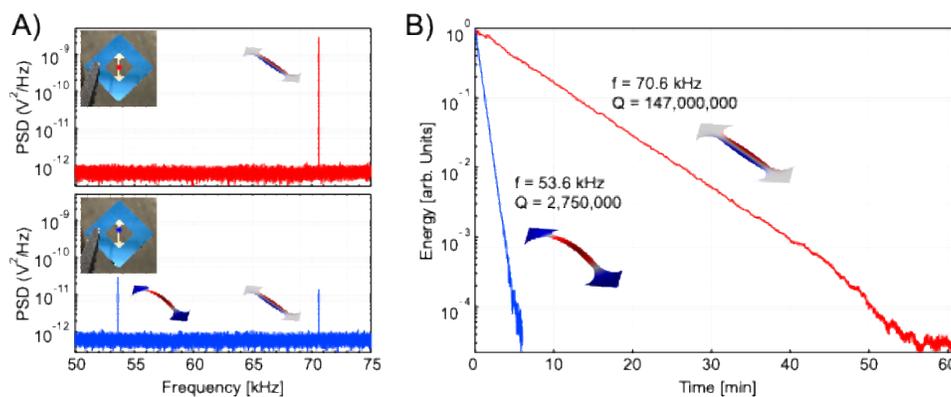

**Figure 5: Nanoribbon Characterization.** The ribbon from Figure 4 is probed with an optical lever producing a) two separate spectra corresponding to two separate probe locations. After locating the two modes, we perform an energy ringdown in b) and extract the Q from an exponential fit.

## DISCUSSION:

One thing that is evident from the data in Figure 5 is that, while the Q of the torsional mode matches the simulated prediction well, the flexural mode does not. This does not fault the quality of the simulation but rather is a result of neglecting extrinsic loss mechanisms such as substrate mode coupling [49] and clamping loss. The dissipation dilution model used in the protocol accounts for all intrinsic loss mechanisms and, therefore, sets an upper bound on the performance of a prestressed nanomechanical resonator.

Another quirk that must be accounted for is the dependence of KOH etch rate on the geometry being released. While KOH reacts with silicon relatively fast, silicon nitride acts as a hard mask. Naturally, large-area devices take significantly longer to etch than their smaller counterparts. Additionally, the KOH etch is anisotropic along the <100> crystal plane, forming pyramids at 54.74° angles [2]. While most geometric features can be etched this way, clamps and other similarly blocked regions near the chip's edge cannot. Thus, fillets in the corner or along the edge of the chip limit the etch rate.

A major limitation of the releasing step is the survivability of the designs being fabricated. While virtually any arbitrary geometry can be patterned into the silicon nitride following sections 1 and 2, the variation in etch rate with geometry also means different parts of a resonator release at different times. This uneven etching gives rise to wild deviations in the stress profile predicted by COMSOL in the KOH bath, adding additional challenges to releasing more extreme resonators. There is no easy way to ensure the survivability of a complex geometry other than through trial and error but maintaining a maximum von misses stress that's far below the yield strength may significantly help.

If the devices are meant to have no material behind the top side film, then the release can be greatly accelerated using a deep silicon etch to excavate most of the silicon from the back of the chip. The remaining silicon can then be removed by a KOH etch as detailed above. Deep reactive ion etching (DRIE) also has the advantage of being done at the wafer or the chip scale, allowing for faster prototyping and characterization of novel devices. Removing excess silicon greatly reduces the time spent in KOH, possibly increasing the yield.

The process laid out in this text has enabled the rapid development of many novel studies and serves as a springboard for more advanced fabrication. Looking ahead, this procedure can be expanded for the creation of double membrane devices [20,48], integrated photonic crystal reflectors [44,45], and extreme geometries from taking a inverse design-based approach to making nanomechanical resonators [46,47].

## ACKNOWLEDGMENTS:

A.R.A. acknowledges support from a CNRS-UArizona iGlobes fellowship; and A.D.H. acknowledges support from a Friends of Tucson Optics Endowed Scholarship. Finally, the reactive ion etcher used for this study was funded by an NSF MRI grant, ECCS-1725571. The protocol was initially developed by A.R.A. and was later modified by A.D.H, C.A.C., and O.A.F. to optimize device fabrication. A.D.H. and D.J.W. co-wrote the manuscript with assistance from all co-authors.